\documentstyle[11pt,paspconf,epsfig]{article}

\begin{document}

\title{An asymmetric relativistic model for FRII radio sources}

\author{T. G. Arshakian\altaffilmark{1} and M. S. Longair}
\affil{Cavendish Laboratory, Madingley Road, Cambridge, CB3 0HE}

\altaffiltext{1}{On leave from Byurakan Astrophysical Observatory, Byurakan 378433, Armenia.} 

\begin{abstract}
An asymmetric relativistic model for FRII radio sources is described which takes account of both relativistic effects and intrinsic/ environmental asymmetries to explain the observed structural asymmetry of their radio lobes. A key feature of the model is jet-sidedness, which can now be determined for about 80\% of the FRII sources in the 3CRR complete sample. It is shown that a simple asymmetric relativistic model can account for a wide range of observational data, and that the relativistic and intrinsic asymmetry effects are of comparable importance. The mean translational speed of the lobes is $\overline{v}_{\rm lobe}=(0.12 \pm 0.04)\,c$. The results are in agreement with an orientation-based unified scheme in which the critical angle separating the radio galaxies from the radio quasars is about $50\deg$.

\end{abstract}


\section{The problems with symmetric relativistic models}

In the simplest kinematic model of FRII radio sources (Ryle \& Longair 1967), the two components move out 
symmetrically from the central active galactic nucleus at the same velocity. Structural asymmetries of the core--hotspot angular distances are attributed to the difference in light travel time from each hotspot to the observer. Many studies have been made of the probability distribution of the velocities of the radio source components from the observed distribution of the ratio of core--hotspot distances (Longair \& Riley 1979; Katgert-Meikelijn {\it et al.} 1980; Banhatti 1980; Best {\it et al.} 1995).  The mean velocity of advance of the hotspots was found to be $\geq 0.2c$, with a considerable spread about the mean velocity, values greater than $0.4c$ being found.

The simple model makes a number of testable predictions about the structure of the sources.  For example, the lobe approaching the observer should be longer than the receding lobe. Many of these sources are now known to contain one-sided radio jets emanating from the nucleus and so, if the cause of this asymmetry is relativistic beaming, the lobe on the jet side should be longer than the counterjet lobe.   Saikia (1984) showed that, in about half of the 36 quasars in his sample, the jet was on the shorter side. The same effect was also present in the smaller quasar samples selected by Bridle (1994) and Scheuer (1995). The latest jet detections in high-resolution VLA observations indicate that there is no tendency for the brighter, or only, jet to lie in the longer lobe for radio galaxies (Hardcastle {\it et al.} 1997). The statistics of 103 FRII radio sources in the 3CRR complete sample (Laing, Riley \& Longair 1983) show that, in about one third of the sources (28 radio galaxies and 8 radio quasars), the jet is on the short side. And so, \emph{the approaching side is not always the longer side}. A further problem arises if the mean space velocity, estimated in a symmetric model to be $\overline{v}_{\rm sm}=(0.27\pm 0.16)\,c$, is compared with that determined  from spectral ageing arguments, $\overline{v}_{\rm sa}=(0.13\pm0.08)\,c$ (Alexander \& Leahy 1987; Liu et al. 1992). This discrepancy suggests that the velocities of lobes may have been overestimated. 

It is not at all unexpected that there should be problems with the simple symmetric model.  For example, McCarthy {\it et al.} (1991) concluded that the one-sidedness of the optical emission-line regions in the vicinity of the radio galaxy provides evidence that \emph{environmental effects contribute to the structural asymmetries of the radio sources}.  As noted by Best {\it et al.} (1995), it is probable that both relativistic and intrinsic asymmetries contribute to the observed properties of the FRII sources. 
       
\section{An asymmetric relativistic model}

\begin{figure}
\begin{center}
\epsfig{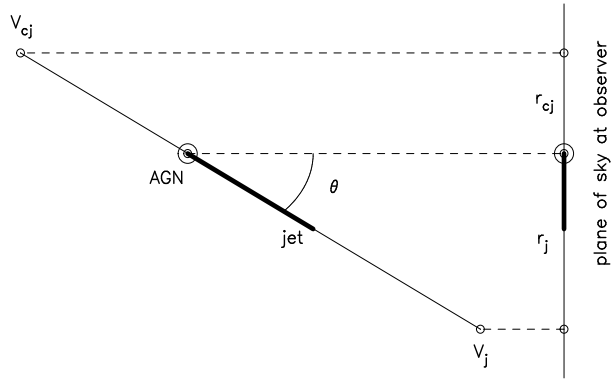}
\caption{The asymmetric relativistic model of FRII radio sources.} \label{fig-1}
\end{center}
\end{figure}

To make quantitative estimates of the relative contributions of the relativistic and intrinsic/environmental effects, we describe an \emph{asymmetric relativistic model}, in which intrinsically asymmetric jets advance through a clumpy asymmetric environment at an angle $\theta$ to the line of sight (Figure~\ref{fig-1}). The asymmetries associated with both the environment and the intrinsic properties of the jets can be described by assuming that the velocity of the lobe in the jet direction $v_{\rm j}$ and that in the counterjet direction $v_{\rm cj}$ are different. We define the structural asymmetry of radio lobes as the \emph{fractional separation difference} $x \equiv (r_{\rm j} - r_{\rm cj})/(r_{\rm j} + r_{\rm cj})$, which is
\begin{equation}
x = \frac{v_{\rm j}-v_{\rm cj}}{v_{\rm j}+v_{\rm cj}}+\frac{2}{c}\,\frac{v_{\rm j}\,v_{\rm cj}}{v_{\rm j}+v_{\rm cj}}\,\cos\theta.
\end{equation}
In the symmetric relativistic model, where $v_{\rm j} = v_{\rm cj}$, the value of $x$ is always positive, but in the asymmetric relativistic model, negative values can be found if the first term on the right-hand side of (1) is negative and of greater magnitude than the second.  By inspection of (1), we see that qualitatively we expect: (i) the radio axis of FRII sources with a jet on the short lobe side ($x \leq 0$, hereafter $-$FRII sources) must lie close to the plane of sky, whilst FRII sources with a jet on the long side ($x \geq 0$, hereafter +FRII) can be observed at all angles; (ii) large positive values of $x$ are found if the lobes with the highest speeds expand in a direction close to the line of sight; small negative values of $x$ are found when the lobes with the lowest speeds expand close to the plane of the sky; (iii) in the case of the symmetric relativistic model, all the sources would be +FRII source, whereas in the purely asymmetric model, with no relativistic effect, there would be equal numbers of +FRII and $-$FRII sources. Thus, we can define an {\it asymmetry parameter} 
\begin{equation}
\varepsilon=1-2\,\,\frac{\mbox{Number}(-\mbox{FRII)}}{\mbox{Number(+FRII)}},
\end{equation}         
which has the property that the relativistic effect is more important than the intrinsic/environmental effect if $0<\varepsilon\leq1$, and {\it vice versa}, if $-1\leq\varepsilon<0$. If $\varepsilon \sim 0$, the contribution of each effect is of comparable importance.

Furthemore, for relativistic sources we would expect that: (i) the mean linear size of $-$FRII sources should be greater than that of +FRII sources; (ii) there should be more two-sided jets in $-$FRII sources than in +FRII sources and more one-sided jets in +FRII sources than in $-$FRII sources; (iii) in orientiation-based unified schemes, intrinsic asymmetries should be more significant for radio galaxies, while relativistic effects should be more important for quasars, that is, $\varepsilon_{\rm G}<\varepsilon_{\rm Q}$. 

\section{The properties of the FRII sources in the 3CRR sample}

Of the radio sources in the complete 3CRR sample, jet-sidedness can be determined reasonably unambigiously for 103 FRII sources, values of $x$ being found for the 71 FRII radio galaxies and 32 FRII radio quasars (for details of the procedures for estimating jet-sidedness, see Arshakian and Longair (1999)).  The \emph{asymmetry parameter} estimated for the joint sample has mean value $\varepsilon_{\rm G+Q}=-0.07\pm0.02$, showing that relativistic effects and intrinsic/environmental asymmetry are of roughly equal importance in determining the structural asymmetries of the radio sources. The intrinsic asymmetry is more important for the radio galaxies, for which $\varepsilon_{\rm G}=-0.3$, while relativistic effects are clearly more important for quasars for which $\varepsilon_{\rm Q}=0.33$.  These results are in accord with the expectations of the asymmetric relativistic model.

The statistics of \emph{one- and two-sided jets} are also consistent with the expectation of the model, namely that the percentage of two-sided jets is greater for $-$FRII (22\%) than for +FRII radio galaxies (19\%) and similarly for $-$FRII (50\%) and +FRII radio quasars (13\%). Since all the sources in the sample have jets, the reverse is true for one-sided jets: there are fewer one-sided jets in $-$FRII as compared with +FRII sources for both radio galaxies and quasars. 

The \emph{mean linear sizes} of 19 low luminosity ($P_{178}<10^{27.7}\,{\rm W}{\rm Hz}^{-1}$) and 81 high luminosity ($P_{178}>10^{27.7}\,{\rm W}{\rm Hz}^{-1}$) sources have also been studied.  The expectation that the $-$FRII sources should be larger than the +FRII sources is found to be the case for the high-luminosity sources (Table~\ref{tbl-1}). No such effect is seen for the low-luminosity sources. Furthermore, the differences in the mean linear sizes of $-$FRII and +FRII radio galaxies and radio quasars can be explained, if the latter are on average observed at smaller angles to the line of sight than the radio galaxies.  

\begin{table}
\caption{The mean projected linear sizes of the sources in the sample with $P_{178}>10^{27.7}\,{\rm W}{\rm Hz}^{-1}$. The numbers of sources are in brackets.} \label{tbl-1} 
\begin{center}\scriptsize 
\begin{tabular}{cll}  
{FRII type} &
\multicolumn{2}{c}{{Mean linear size} (kpc)}       \\ 
RG/RQ  & $-$FRII & $+$FRII             \\ \hline \hline \\
RG+RQ  & 290  (27)                    & 220  (54)        \\ 
RG      & 290  (19)                    & 280  (30)           \\ 
RQ     & 280   (8)                    & 140  (24)           \\  
\end{tabular}
\end{center}
\end{table}

\section{Expansion speeds and the unified scheme}

In order to quantify the relative importance of intrinsic/environmental asymmetries as opposed to relativistic asymmetries, we can write the velocities $v_{\rm j}$ and $v_{\rm cj}$ as $v_{\rm j}=v_0 + v_{\rm d}$ and $v_{\rm cj} = v_0 + v_{\rm cd}$, where $v_0$ is the average velocity of lobes and $v_{\rm d}$ and $v_{\rm cd}$ are caused by intrinsic/environmental asymmetries on the jet and counterjet sides respectively. For the sake of illustration, we assume that the distribution function of $v_0$ is of gaussian form and that the dispersion in the intrinsic/environmental velocity dispersion is another independent gaussian. It is then straightforward to work out the expected distributions of $x$, selecting mean speeds and velocity dispersions to agree with the observed value $\varepsilon_{\rm G+Q}$$=-0.07$. For the joint sample of radio galaxies and quasars, we find that the mean expansion speed of the lobes and its standard deviation are $\overline{v}_{\rm lobe} \sim (0.12 \pm 0.043)\,c$, which is similar to the results of spectral ageing analyses, ($0.13 \pm 0.08)\,c$, significantly less than the mean speed estimated from the symmetric model. As noted above, the mean speeds are overestimated in the symmetric model because structural asymmetries are attributed entirely to the light travel times, whereas our improved analysis shows that, in fact, both relativistic effects and intrinsic asymmetry contribute equally to the structural asymmetry.

In orientation-based unified schemes for radio galaxies and quasars (Barthel 1989), these are the same class of object but  viewed at different angles to the line of sight. A critical angle $\theta_{\rm c} \sim 45{\deg}$ separates radio galaxies from the radio quasars. By modelling the observed distribution function of apparent velocities for different critical angles, the predicted asymmetry parameter for both radio galaxies and quasars can be estimated (Table~\ref{tbl-2}). The observed values are $\varepsilon_{\rm G}=-0.3$ and $\varepsilon_{\rm Q}=0.33$. Excellent agreement between the predicted and observed asymmetry parameters is found for $\theta_{\rm c} \sim 50{\deg}$ as can be seen from Table~\ref{tbl-2}. We conclude that the asymmetric relativistic model is in satisfactory agreement with an orientation-based unified scheme with  $\theta_{\rm c} \sim 50{\deg}$.

\begin{table}
\caption{Predicted asymmetry parameters for radio galaxies and quasars according to an orientation-based unified scheme for the asymmetric relativistic model.} \label{tbl-2}
\begin{center}\scriptsize
\begin{tabular}{ccc|ccc}    
$\theta_{\rm c}^{\circ}$ & $\varepsilon_{\rm\small{G}}$ & $\varepsilon_{\rm\small{Q}}$ & $\theta_{\rm c}^{\circ}$ & $\varepsilon_{\rm\small{G}}$ & $\varepsilon_{\rm\small{Q}}$ \\ \hline \hline
10 & $-0.08$ & 0.45 & 50 & $-0.27$ & 0.33 \\
20 & $-0.11$ & 0.39 & 60 & $-0.35$ & 0.27 \\
30 & $-0.12$ & 0.36 & 70 & $-0.46$ & 0.17 \\
40 & $-0.16$ & 0.32 & 80 & $-0.81$ & 0.06 

\end{tabular}
\end{center}
\end{table}

\section{Conclusions}
\begin{itemize}
\item The proposed asymmetric relativistic model of FRII radio sources can take account of both relativistic and intrinsic/environmental asymmetry effects.  The predictions of the model are in agreement with observational data, and the observational data indicate that both effects are of comparable importance.
\item The mean expansion speed of lobes is about $(0.12 \pm 0.043)\,c$, a value consistent with values found from synchrotron ageing arguments.
\item The data for radio galaxies and quasars are consistent with orientation-based unification schemes in which the   critical angle separating the two types of object is $\theta_{\rm c} \approx 50{\deg}$. 
\end{itemize}

\acknowledgments

We are very grateful to Julia Riley for helpful comments and kindly supplying up-to-date information about the 3CRR sample, and to Andrew Blain for his critical reading of drafts of this paper.




\begin{references}
\reference Alexander, P., Leahy J.\,P., 1987, \mnras, 225, 1
\reference Arshakian, T., Longair, M.S., 1999. \mnras, (in preparation)
\reference Banhatti, D.\,G., 1980, \aap, 84, 112
\reference Barthel, P.\,D., 1989, \apj, 336, 606
\reference Best, P.\,N., Bailer, D.\,M., Longair, M.\,S., Riley, J.\,M., 1995, \mnras, 275, 1171
\reference Bridle, A.\,H., et al., 1994, \aj, 108, 766 
\reference Hardcastle, M.\,J., et al., 1997, \mnras, 288, 859 
\reference Katgert-Meikelijn, J., Levi, C., Padrielli, L., 1980, \aaps, 40, 91
\reference Laing, R.\,A., Riley, J.\,M., Longair, M.\,S., 1983, \mnras, 204, 151
\reference Longair, M.\,S., Riley, J.\,M., 1979, \mnras, 188, 625
\reference Lui, R., Pooley, G., Riley, J.\,M., 1992, \mnras, 257, 545
\reference McCarthy, P.\,J., van Breguel, W.\,J.\,M., Kapahi, V.\,K., 1991, \apj, 371, 478
\reference Ryle, M., Longair, M.\,S., 1967, \mnras, 136, 123 
\reference Saikia, D.\,J., 1984, \mnras, 209, 525 
\reference Scheuer, P.\,A.\,G., 1995, \mnras, 277, 331
\end{references}
\end{document}